\documentstyle[11pt,newpasp,twoside,epsf]{article}
\begin{document}
\title{Bump Cepheids and the Stellar Mass-Luminosity Relation}
 \author{Stefan C.\ Keller}
\affil{Institute of Geophysics and Planetary Physics, LLNL, \\ Livermore, CA 94550, U.S.A.}

\author{Peter R.\ Wood}
\affil{Research School of Astronomy and Astrophysics, \\ Australian National University, A.C.T.~2600, Australia}

\begin{abstract}
We present the results of non-linear pulsation modeling of bump
Cepheids in the LMC and SMC. By obtaining an optimal fit to the
observed MACHO $V$ and $R$ lightcurves we can determine the
fundamental parameters of each Cepheid, namely mass, luminosity,
effective temperature, distance modulus and reddening. We are able to
describe the mass-luminosity relation for core-He burning intermediate
mass stars. The mass-luminosity relation depends critically upon the
level of internal mixing during the course of the star's main-sequence
evolution. Under the paradigm of convective core overshoot, our
results enable us to place tight quantitative limits of the level of
overshoot. We derive an overshooting parameter of $\Lambda_c$ of
0.65$\pm$0.03 and 0.67$\pm$0.04~$l/H_p$ for the LMC and SMC
respectively.
\end{abstract}

\section{Introduction}

Cepheids form the first step to extra-galactic distances and are
fundamental to modern observational cosmology. It remains a goal to
have theoretical models capable of predicting the period-luminosity
relation and its metallicity dependence. The regularity of Cepheid
pulsation provides a set of well defined observational parameters with
which to confront the predictions of theoretical models of stellar
pulsation. In this way, Cepheids provide close scrutiny of the
accuracy of input physics within pulsation models.

One of the weakest points in our description of the internal structure
of intermediate to massive stars is the description of convection in
the vicinity of the convective core. Ongoing debate centers of the
degree of extension of the convective core above its classical
Schwarzschild boundary. Convection is of course a fundamentally 3d
problem awaiting hydrodynamical simulation such as is planned with
Djehuty (see Bazan in these proceedings).

For the time being it suffices to treat the extension of the classical
convective core by means of mixing-length theory. The convective core
overshoot parameter $\Lambda_c$ sets the height (as a fraction of the
pressure scale height) to which convective motion extends the core
into the formally convectively stable region.

The size of the convective core determines the helium core mass laid
down by the end of core H burning. A larger He core mass increases the
luminosity of the post-main-sequence evolution and hastens its
pace. Consequently the mass-luminosity relation for Cepheids is
determined by the level of convective core overshoot whilst the star
was on the main-sequence.

\section{The Modeling Procedure}

The lightcurve morphology of a Cepheid is determined by the stellar
mass, luminosity and effective temperature. A feature of the
lightcurves of a subset of the Cepheid population is the presence of a
bump either preceding or following maximum light. This is the bump in
bump Cepheids (see Figure~1). The bump arises from the
2:1 resonance between the fundamental mode and the second overtone. It
becomes particularly prominent when the period ratio of these two
modes ($P_{02}$) is close to two.

We have selected a sample of bump Cepheids from the MACHO photometric
database in both the LMC (20 stars) and the SMC (10 stars). The
details of the non-linear pulsation code are provided in Keller \&
Wood (2002). In contrast to the work of Bono, Castellani \& Marconi
(2002) we use only stellar pulsation and atmosphere theory - we do not
make recourse to the assumption of existing mass-luminosity relations.

Throughout standard abundances for LMC and SMC populations are assumed
(Z=0.008 and 0.004 respectively). This us with three fundamental parameters
to describe the pulsation envelope: M, L and $T_{\rm eff}$. Hence we
require three constraints to determine these. The first constraint is
that the fundamental period of the model matches that observed and
this is achieved from linear theory. The second and third are obtained
from the lightcurve fit from the nonlinear code. This procedure is
illustrated in Figure~1.

In Figure 1 effective temperature is varied vertically and P$_{02}$
horizontally. As P$_{02}$ is varied the phase of the bump is modified;
$T_{\rm eff}$ varies the amplitude. The best fit to the observed
lightcurve is located in the central panel. In this way we have
determined the fundamental parameters of the Cepheid: M, L, $T_{\rm eff}$,
distance and reddening.

\begin{figure}
\plotone{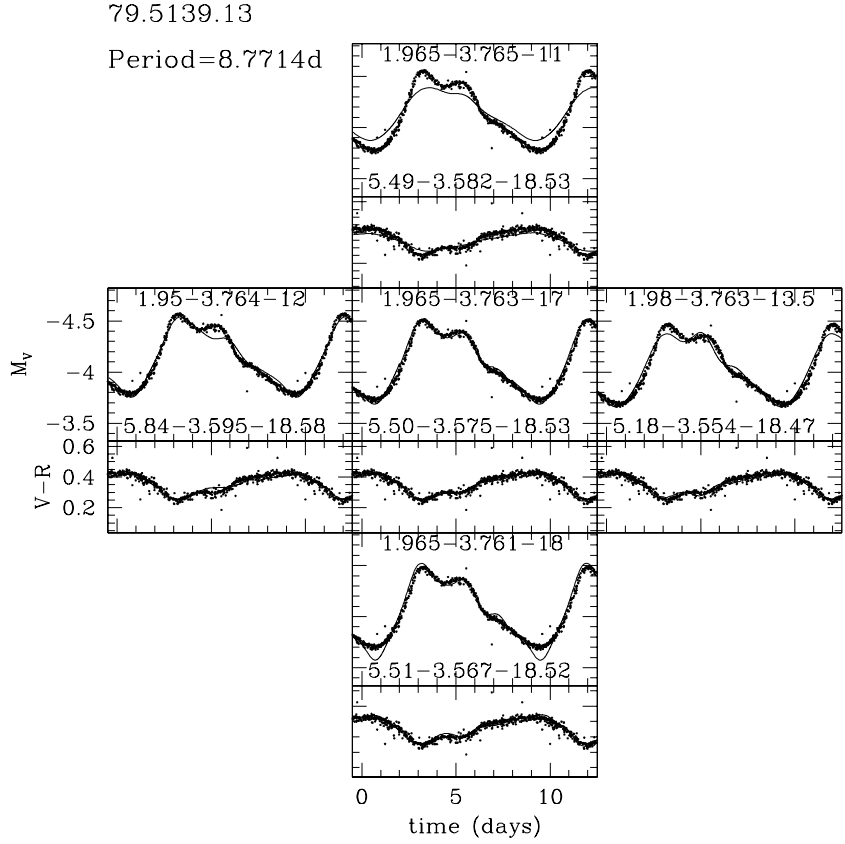}
\caption{Constraining the mass and luminosity of the bump Cepheid: M$_V$ and
$V$$-$$R$ against time for five models of the MACHO Cepheid
79.5139.13. Lines show model output except that the observed $V$ and
$V$$-$$R$ have been shifted vertically to give the best match to the
model in each case. This provide the distance modulus and
reddening. The parameters shown in each of the five boxes are, top
line: $P_{02}$, $T_{\rm eff}$ and the initial perturbation velocity
applied to the envelope, bottom line: mass, luminosity and distance
modulus. }

\label{fig.5panel}
\end{figure}

\section {Caveat Pulsator}

The above models treat convective energy transport via the
mixing-length approximation. This approximation is expected to break
down at cooler temperatures as the convective zone becomes a
substantial fraction of the envelope. A consequence is that our models
do not reproduce the red edge of the instability strip.  In the
vicinity of the red edge the pulsation amplitude is too high. The
additional dissipative effect of turbulent convection is present in a
real Cepheid atmosphere.

To avoid this shortcoming we have selected our sample of bump
Cepheids close to the blue edge of the instability strip. This region
has the added advantage that driving and hence amplitude, is very
sensitive to effective temperature.

\begin{figure}
\plotfiddle{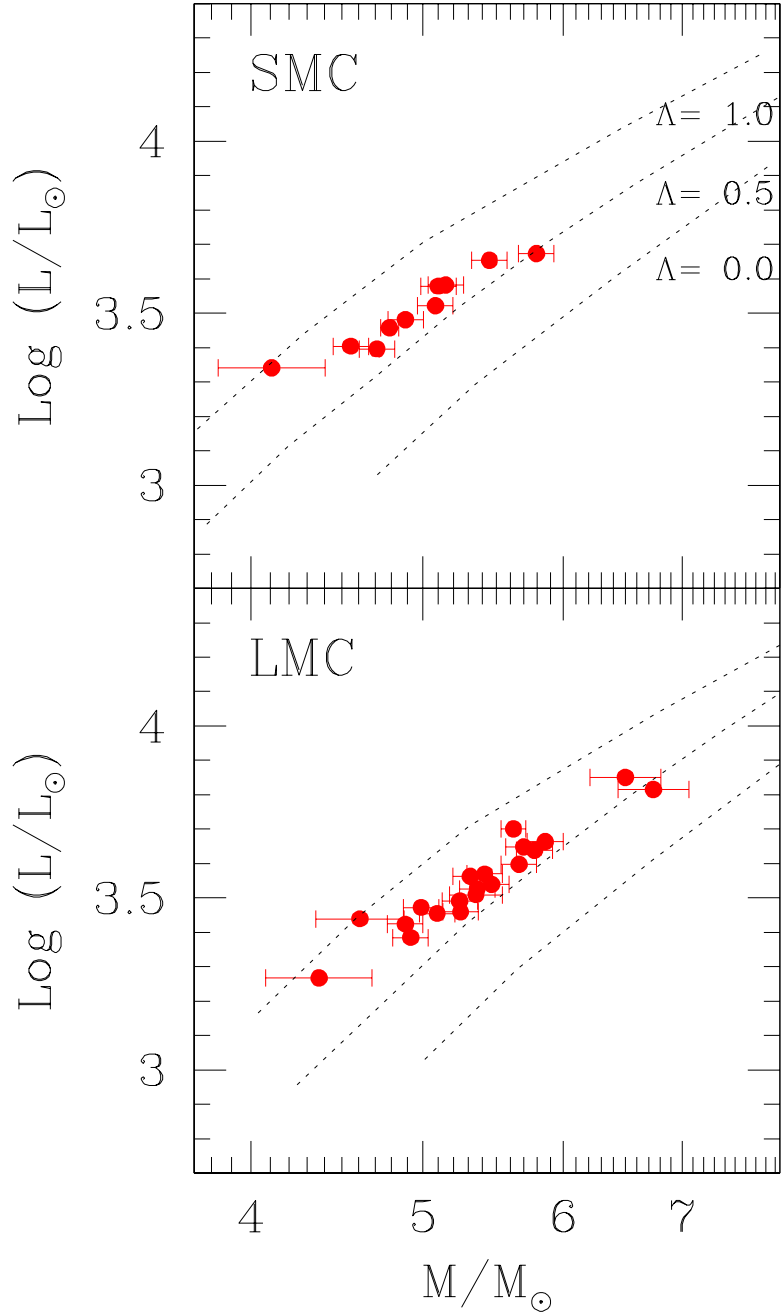}{18.15truecm}{0}{105}{105}{-300}{-150}

\caption{The mass-luminosity relation for the sample of 10 SMC (top) and 20 LMC (bottom) bump Cepheids. Overlaid are mass-luminosity relations from evolutionary models with three values of the convective core overshoot parameter ($\Lambda_c$=0.0, 0.5 and 1.0~$l/H_p$). }
\end{figure}

\section{The Mass-Luminosity Relation}

Our results for our sample of LMC and SMC bump Cepheids are presented
in Figure 2. Overlaid are M-L relations from stellar evolutionary models
for three values of $\Lambda_c$. It is evident that both samples are
significantly mode luminous than classical models. This difference
amounts to a reduction of 20\% in mass from classical models, with an
optimal $\Lambda_c$ of 0.65$\pm$0.03 pressure scale heights for the
LMC sample and 0.67$\pm$0.04 for the SMC. We also note that we do not
see evidence for a metallicity dependence in the level of convective
core overshoot.

This runs counter to other circumstantial evidence that would suggest
that core overshoot should increase as we decrease metallicity. The
findings of Venn (1999 and 1995) indicate a higher level of chemical
enrichment amongst A supergiants, more massive cousins of the Cepheids
examined here. Rotationally induced mixing has been proposed as the
mechanism for this more efficient mixing in the stellar envelope. This
suggests a generally more rapid rotation amongst lower metallicity
stars (Keller et al.~2001b). If rotation were responsible we would
expect the SMC stars to be of generally higher luminosity relative to
classical models.

\section{Summary}
The problem of reconciling pulsation masses with evolutionary masses
for Cep\-heids has a long history much of which has been resolved with
the introduction of the OPAL opacities. The discrepancy between
pulsation and classical evolution mass that we have demonstrated here
is the remainder of the debate which has not been removed with
improved input physics.

Rather, the discrepancy shown here has a basis in a higher level of
internal mixing within intermediate mass main-sequence stars. Our
findings are supported by a number of studies using linear pulsation
analysis (Sebo \& Wood 1995) and studies of stellar populations
(Keller et al.~2001a) which show the need for a level of convective
core overshoot of order 0.5~$l/H_p$. Our study of bump Cepheids has
enabled us to place stringent limits on $\Lambda_c$ and we now wait to
see if the results of 3d hydrodynamical simulations such as those
available from Djehuty match that observed.

\acknowledgements
Work performed by SCK was performed under the auspices of the U.S.~Department of Energy, National Nuclear Security Administration by the University of California, Lawrence Livermore National Laboratory under contract No. W-7405-Eng-47.

\vfill\eject

\begin{references}
\reference Bono, G., Castellani, V., \& Marconi, M.\ 2002, ApJL, 565, L83
\reference Keller, S.~C., Da Costa, G.~S., \&  Bessell, M.~S. 2001a, AJ, 121, 905
\reference Keller, S.~C., Grebel, E.~K., Miller, G.~J., \& Yoss, K.~M.\ 2001b, AJ, 122, 248
\reference Keller, S.~C., \& Wood, P.~R. 2002, ApJ, accepted, astro-ph/0205555
\reference Sebo, K.~M., \& Wood, P.~R. 1995, ApJ, 449, 164
\end{references}
\end{document}